\documentclass[11pt, DIV10,a4paper]{article}
\usepackage{color}
\usepackage{float}
\usepackage[bf,small]{caption2}
\usepackage{comment}
\usepackage{amsmath}
\usepackage{bigints}
\usepackage{amsopn}
\usepackage{amsfonts}
\usepackage{amsthm}
\usepackage{amssymb}
\usepackage{amsbsy}
\usepackage{tocbibind}
\usepackage[a4paper,colorlinks,breaklinks,unicode]{hyperref}%backref,
\usepackage{makeidx}
\usepackage{pst-all}
\usepackage{epsfig,psfrag}
\usepackage{graphicx}
\usepackage{amsmath}
\usepackage{epstopdf}
\usepackage{tabularx}
\usepackage{subfigure}
\usepackage{setspace}
\usepackage[mathscr]{euscript}
\usepackage[margin=1in]{geometry}
\usepackage{multirow}
\usepackage{slashbox}
\newcommand {\ctn}{\cite}

\usepackage{url}
\newtheorem{lemma}{Lemma}[section]
\newtheorem{proposition}{Proposition}[section]

\newtheorem{result}{Result}[section]
\newtheorem{algo}{Algorithm}[section]

\normalsize
\begin{document}

\title{\textbf{On Single Variable Transformation Approach to Markov Chain Monte Carlo}}
\author{ Kushal K. Dey$^{\dag +}$ , Sourabh Bhattacharya$^{*}$ }
\date{}
\maketitle
\begin{center}
$^{\dag} $    University of Chicago, IL  \\
$^{*}$   Indian Statistical Institute, Kolkata \\
$+$ Corresponding author:  \href{mailto: kkdey@uchicago.edu}{kkdey@uchicago.edu}
%,  \href{mailto: bhsourabh@gmail.com}{bhsourabh@gmail.com}\\
\end{center}

\section{Introduction}

In today's times, Markov Chain Monte Carlo (MCMC) methods have everyday use in Statistics and other disciplines like 
Computer Science, Systems Biology and Astronomy.  This technique of generating random samples even 
from very high dimensional spaces involving very complicated data likelihoods and posterior distributions 
has simplified many pressing real life problems in recent times. In particular, Bayesian computation, 
simulation from complex posterior distribution and asymptotics of Bayesian algorithms have benefited 
a lot from this mechanism (see Gelfand and Smith \cite{Gelfand1990}, Tierney \cite{Tierney1994}, 
Gilks \emph{et al} \cite{Gilks1996}). 
A very standard approach of simulating from multivariate distributions is to use the Metropolis-Hastings (MH) algorithm 
\cite{Hastings}\cite{Metropolis} using the random walk proposal. We refer to such algorithm as
the Random Walk Metropolis Hastings (RWMH) algorithm. The convergence and optimal scaling of this 
algorithm has been extensively studied \cite{Gelman}. 
%However, there are obvious scopes for improving 
%upon this algorithm, pertaining mainly to the choice of proper proposal distribution and 
%the time-complexity associated with the process. 
However, despite the advances, there are certain glaring problems that one may encounter while using RWMH. 
For very high dimensional, non-standard target distributions, choosing the scales optimally is not feasible in practice,
and hence, attempts of jointly updating the parameters using RWMH face serious drop in the acceptance rate, which,
in turn, leads to poor convergence. 
Methods of adpatively selecting the scales usually take very large number of iterations to even converge
to the optimal scales; particularly in complex and very high-dimensional situations, this exercise is computationally
burdensome in the extreme. The alternative method of updating the parameters sequentially is not only
computationally burdensome in high-dimensional problems, high posterior correlation among the parameters
usually cause very slow convergence. These issues are discussed in much detail in \cite{Dey2013(2)}. 
%This issue 
%convergence of RWMH to the target density is pretty slow and one requires too many iterations, 
%largely due to the fact that in RWMH, we need to update each co-ordinate at a time and this may lead 
%to very small acceptance probability for high dimensions. 

The TMCMC methodology proposed in Dutta and Bhattacharya \cite{Dutta2011} tries to address these problems. 
The methodology uses simple deterministic transformations using (typically) a single random variable  
having an appropriately chosen proposal density. In this paper, we primarily study one version, 
termed as the Additive TMCMC (ATMCMC) method, and deal with the ergodic behavior of the chain 
in high dimensions. Our aim is to present a comparative study of ATMCMC and 
the standard RWMH algorithm with respect to their ergodic behaviors. \\ [3 pt]

This paper is organized as follows. In \textbf{Section 2}, we present the ATMCMC 
algorithm and discuss the intuition behind this algorithm. In \textbf{Section 3}, we discuss 
some theoretical results regarding the ergodic behavior of the chain. \textbf{Section 4} focuses 
on how to optimally select the proposal density for the chain when the target density has 
a product structure. In \textbf{Section 5}, we present the comparative simulation study of ATMCMC 
and RWMH and analyze the results. 

\section{Algorithm}

We first briefly describe how additive TMCMC (ATMCMC) works. We explain it for the bivariate case 
\--- the multivariate extension would analogously follow. Suppose we start at a point 
$(x_{1},x_{2})$. We generate an $\epsilon >0$ from some pre-specified proposal distribution 
$q$ defined on $\mathbb{R}^{+}$. Then in additive TMCMC we have the following four possible "move-types": 
\begin{eqnarray}
(x_{1},x_{2}) \rightarrow (x_{1}+\epsilon,x_{2}+\epsilon) \nonumber \\
(x_{1},x_{2}) \rightarrow (x_{1}+\epsilon,x_{2}-\epsilon) \nonumber \\
(x_{1},x_{2}) \rightarrow (x_{1}-\epsilon,x_{2}+\epsilon) \nonumber \\
(x_{1},x_{2}) \rightarrow (x_{1}-\epsilon,x_{2}-\epsilon) \nonumber \\
\end{eqnarray} 
This means we are moving along two lines in each transition from the point $(x_{1},x_{2})$, 
one parallel to the line ${y=x}$ and the other parallel to the direction ${y=-x}$. Each of the four 
transitions described above are indexed as $I_{k}$ for the $k$th transition, where $k$ varies 
from 1 to 4 in the bivariate case, and in general from 1 to $2^{d}$ in $\mathbb{R}^{d}$. 
For simplicity we assume that the move-types are chosen with equal probability; see 
Dutta and Bhattacharya \cite{Dutta2011} for the
general case. As with the standard RWMH case, we do attach some probabilities with accepting/rejecting 
the proposed move such that the reversibility condition is satisfied thereby guaranteeing convergence. 
Formally, the algorithm may be presented as follows.

\newcommand{\topline}{\hrule height 1pt width \textwidth \vspace*{2pt}}
\newcommand{\botline}{\vspace*{2pt}\hrule height 1pt width \textwidth \vspace*{4pt}}

\newcommand{\bx}{\textbf{x}}
\newcommand{\by}{\textbf{y}}

\begin{algo}\label{algo:algo1}
Suppose we are at $\bx_{n}=(x_{1},x_{2},\cdots,x_{d})$ at the $n$th iteration. 

\begin{enumerate}

\item Generate $\epsilon \sim g(\cdot) $ on $\mathbb{R}^{+}$. 
\item Select randomly one move type and define
\[ b_{1},b_{2}, \cdots, b_{d} \stackrel{iid}{\sim}{ Discr Unif \{-1,1 \}} \]
\begin{equation}
\by= (x_{1}+b_{1}\epsilon, x_{2}+ b_{2}\epsilon, \cdots, b_{d}\epsilon)  
\end{equation} 

\begin{equation}
\alpha(\bx,\epsilon)= min \left \{ 1,\frac{\pi(\by)}{\pi(\bx_{n})}\right \} 
\end{equation}

\item  Set 
$\bx_{n+1}=   \left \{ \begin{array}{lll}
\by &  with \hspace{0.3 cm}prob. & \alpha(\bx_{n},\epsilon)  \\
\bx_{n}  & with \hspace{0.3 cm} prob. &  1-\alpha(\bx_{n},\epsilon)  \\
\end{array}     \right  \}$ \\[3.5 pt]

\end{enumerate}
\end{algo}

\begin{figure}%[htp]
\centering
\includegraphics[trim= 2cm 8cm 2cm 8cm, clip=true, width=15cm,height=8cm]{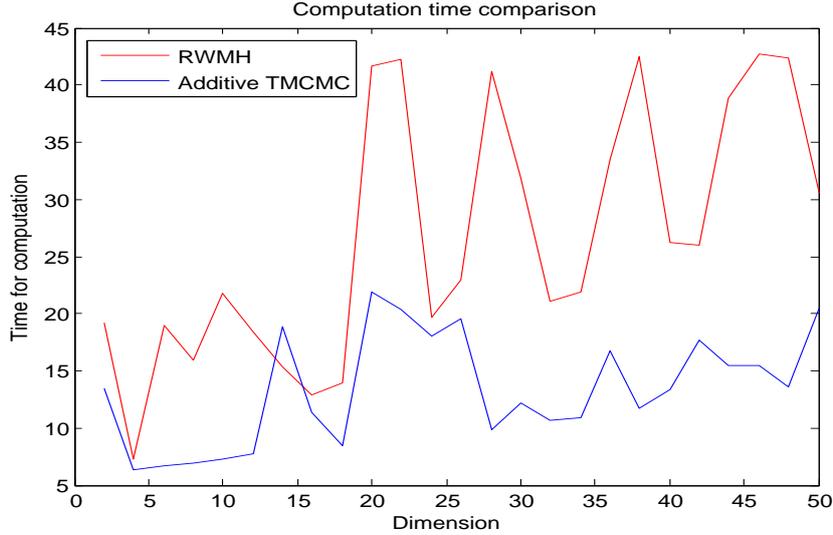}
\caption{\emph{Computation time (in MATLAB R2013b) of one run of 100,000 iterations with RWM and TMCMC 
algorithms corresponding to 
dimensions varying from 2 to 50 with target density being product of $N(0,5)$ and the proposal density 
for additive TMCMC being $TN_{>0}(0,1)$ (truncated $N(0,1)$ left truncated at $0$) and for RWMH proposal, 
every component has $N(0,1)$ distribution. It is observed that TMCMC has consistently less computation 
time compared to RWM, specially for higher dimensions.}}
\label{fig:comptime}
\end{figure}

Now we intuitively discuss why ATMCMC is a better option compared to the RWMH algorithm. 
Firstly, we tested using simulation experiments (all conducted in MATLAB R2013b) that our algorithm requires 
less computational time to run 
compared to RWMH (see \textbf{Fig~\ref{fig:comptime}}).  \\[2 pt]

%But this is not the major issue. 
Secondly, and more importantly, ATMCMC is expected to have much higher acceptance rate than
RWMH. We discuss this as follows.

In a standard RWMH algorithm in $d$ dimensions, 
we need to generate $d$ many $\epsilon_{i}$'s, for $i \in \{1,2,\cdots,d \}$. 
For simpliicty of illustration, assume that the target density $\pi$ is the product density, 
$\pi=\prod_{i=1}^{d}{f()}$ of iid components $f$. 
Then the acceptance rule for RWMH comprises the ratio

\[ \frac{\pi(\bx+\epsilon)}{\pi(x)} = \prod_{i=1}^{d} \frac{f(x_{i}+\epsilon_{i})}{f(x_{i})}. \]

If $d$ is very large, then, by chance, we may obtain some very small or large values of 
$\epsilon_{i} \sim q(\cdot)$ (note that $5 \%$ observations are expected to lie outside the 
$95 \%$ confidence region and these are the points that are problematic). This would result in 
certain very small values of $f(x_{i}+\epsilon_{i})$ for some $i$ and thereby drastically reduce the 
above ratio. So, the chain has the problem of remaining stuck at a point for a long time. Note that 
ATMCMC uses only one $\epsilon$ to update all the co-ordinates using sign change and 
this counters the above problem. So, we can expect a much higher acceptance rate for ATMCMC over the RWMH algorithm. 
But there are two pertinent questions here. Firstly, how much can we 
improve on the RWMH algorithm in terms of the acceptance rate? Secondly, how would the sample we get 
using the ATMCMC method compare to the RWMH algorithm in terms of the convergence of the iterates to 
the target density and the mixing among the iterates once the target is attained? We address the 
first issue in \textbf{Section 4} and the second in \textbf{Section 5}.

\section{Ergodic Properties of ATMCMC}

In case of Markov chains on discrete spaces, there is a well-established notion of irreducibility. 
However, on general state spaces, such a notion no longer works. This is why we define $\psi$ irreducibility. 
A Markov chain is said to be \emph {$\psi$-irreducible} if there exists a measure $\psi$ such that 
\newcommand*{\bigchi}{\mbox{\Large$\chi$}}
\begin{equation}
\psi(A) > 0 \implies \exists n \hspace{0.5 cm} with \hspace{0.5 cm}   P^{n}(x,A) > 0  \hspace{1 cm}  \forall x \in \bigchi
\end{equation}
where $\bigchi$ is the state space of the Markov chain (in our case, it would most often be $\mathbb{R}^{d}$ for some $d$). 
For convergence of the process, we must ensure that it is $\mu$-irreducible, where $\mu$ is the Lebesgue measure. 
We also need additional concepts of aperiodicity and \emph{small} sets. A set $E$ is said to be $small$ if there exists 
$n>0$ , $\delta>0$ and some measure $\nu$ such that 
\begin{equation}\label{eq:small}
P^{n}(x,\cdot) > \delta\nu(\cdot) \hspace{1 cm} x \in E
\end{equation}

A chain is called \emph{aperiodic} if the $g.c.d$ of all such $n$ for \textbf{Eqn~\ref{eq:small}} holds, is 1. 
All these concepts of $\mu$-irreducibility, aperiodicity and small sets are very important for laying the 
basic foundations of stability. The following theorem due to Dutta and Bhattacharya \cite{Dutta2011} 
establishes these properties for the ATMCMC chain. 

\begin{result} \label{Theorem 1}
Let $\pi$ be a continuous target density which is bounded away from 0 on ${\mathbb{R}}^{d}$. 
Also, let the proposal density $q$ be positive on all compact sets on $\mathbb{R}^{+}$. Then, 
every non-empty bounded set in ${\mathbb{R}}^{d}$ is small, and this can be used to show that 
the chain is both $\lambda$-irreducible and aperiodic. 
\end{result}

A proof of this result can be found in Dutta and Bhattacharya \cite{Dutta2011}, along with a 
graphical interpretation; see also Dey and Bhattacharya \cite{Dey2013}. In fact, in Dutta and Bhattacharya \cite{Dutta2011}, 
a stronger result has been proved that for 
any $n>d$ ($d$ represents the dimensionality of the state space), the minorization condition is satisfied.
%
%\[ P^{n}(x,A) \geq \delta\lambda(\cdot) \hspace{1 cm} x \in E \; (bdd.\; Borel \; set) \]
%
%where $\lambda$ is the Lebesgue measure and he explicit form of the 
%$\delta$ depends on the bounds on the proposal density. 
From the monorization condition, $\lambda$ irreducibility follows trivially. Aperiodicity follows 
because the above result is true for all $n>d$ and the $g.c.d$ of such $n$ is $1$. 

Let $P$ be the transition kernel of a $\psi$-irreducible, aperiodic Markov chain with the stationary distribution $\pi$. 
Then the chain is geometrically ergodic if $\exists$ a function $V \geq 1$, which is finite at least one point, 
and also constants $\rho\in (0,1)$ and $M~(<\infty)$, such that

\begin{equation}\label{eq:geo}
||P^{n}(x,\cdot)- \pi(\cdot)||_{TV} \leq MV(x)\rho^{n}  \hspace{0.5 cm} \forall n \geq 1,
\end{equation}
where $||\nu||_{TV}$ denotes the \emph{total variation norm}, defined as
$$ ||\nu||_{TV}= \underset{g: |g| \leq V}{\sup} \nu(g) $$

%The reason for preferring geometric ergodicity is that under this condition, 
Apart from ensuring geometric rate of convergence of the Markov chain, another utility of
geometric ergodicity is that one can apply 
Central Limit Theorem to a wide class of functions of the Markov chain, and hence, one can also investigate 
stability of these ergodic estimates (see Roberts, Gelman and Gilks \cite{Gelman}). A very standard way of 
checking geometric ergodicity is a result that involves the Foster-Lyapunov drift criteria. 
$P$ is said to have a geometric drift to a set $E$ if there is a function $V \geq 1$, finite for at 
least one point and constants $\lambda <1 $ and $c< \infty $ such that

\begin{equation}\label{eq:drift}
PV(x) \leq \lambda V(x) + c 1_{E}(x), 
\end{equation}
where $PV(x)= \int{V(y)P(x,y)dy}$ is the expectation of $V$ after one transition 
given that one starts at the point $x$. Theorems 14.0.1 and 15.0.1 in Meyn and Tweedie \cite{MeynTweedie} 
establish the fact that if $P$ has a geometric drift to a small set $E$, then under certain regularity conditions, 
$P$ is $\pi$-almost everywhere geometric ergodic and the converse is also true.

The first result we present is basically adaptation of a result due to Mengersen and Tweedie \cite{MengersenTweedie}. 
We now show a sufficient condition that would ensure that \textbf{Eqn~\ref{eq:drift}} holds.

\begin{lemma}\label{Lemma 1}
If $\exists$ $V$ such that $V \geq 1$ and finite on bounded support, such that the following hold:
%\begin{equation}\label{eq:1con}
%\underset{|x| \rightarrow \infty}{\lim\sup}~{\frac{PV(x)}{V(x)}}  <  1 
%\end{equation}
%\begin{equation}{\label{eq:infcon}}
%{\frac{PV(x)}{V(x)}} <  \infty  \hspace{1 cm} \forall x
%\end{equation}
\begin{align}
\underset{|x| \rightarrow \infty}{\lim\sup}~{\frac{PV(x)}{V(x)}}  &<  1
\label{eq:1con}\\
\quad\quad {\frac{PV(x)}{V(x)}} &<  \infty  \hspace{1 cm} \forall x.
\label{eq:infcon}
\end{align}
Then this $V$ satisfies the geometric drift condition in \textbf{Eqn~\ref{eq:drift}}, 
and hence the chain must be geometrically ergodic. Also, if for some $V$ finite, 
the geometric drift condition is satisfied, then the above condition must also hold true.
\end{lemma}

\begin{result}\label{theorem:geo_additive}

If $\pi$, the target density, is sub-exponential and has contours that are nowhere 
piecewise parallel to $ \{ x: |x_1|=|x_2|=\cdots=|x_d| \}$, 
then the additive TMCMC chain satisfies geometric drift if and only if

\begin{equation}
\underset{\|x\| \rightarrow \infty}{\lim\inf}~ Q(x, A(x)) > 0,
\label{eq:liminf_Q_additive}
\end{equation}
where $A(x)$ denotes the acceptance region when $x$ is updated, and $Q(x,A(x))$ denotes the
probability of the acceptance region under the ATMCMC proposal distribution associated with the density 
$q(\cdot)$ of $\epsilon$. %See \cite{Dey2013} for the complete mathematical details.

%This result implies that this condition is all we require to show geometric ergodicity for the additive TMCMC chain.
%In this case, one can check that the function $V$ that satisfies the geometric drift condition is 
%$ V(x) = \frac{c}{\sqrt{\pi(x)}}$ corresponding to some $c > 0$. 

\end{result}

A proof of this result is given in Dey and Bhattacharya \cite{Dey2013}. 
A similar result holds true for the RWMH algorithm as well (see Jarner and Hansen \cite{JarnerHansen} 
and Roberts and Tweedie \cite{RobertsTweed}) except that there we do not need the constraint that the 
contours are not piecewise parallel to $ \{ x: |x_1|=|x_2|=\cdots=|x_d| \}$, but this is true for most densities 
we commonly encounter. Even if this condition is not satisfied, we can still show geometric ergodicity 
for a modified TMCMC chain with moves from $(x_1,x_2,\cdots,x_d)$ to 
$(x_1+b_1 c_1 \epsilon_1, x_2+b_2 c_2 \epsilon_2, \cdots, x_d+b_d c_d \epsilon_d)$ 
where $c_i$'s are some positive scalars not all equal. 

\section{Optimal Scaling of Additive TMCMC}

\begin{figure}[h]
\centering
\subfigure [ Small proposal variance sample path]{ \label{fig:fig3}
\includegraphics[trim= 0cm 10cm 0cm 10cm, clip=true, width=7cm,height=3cm]{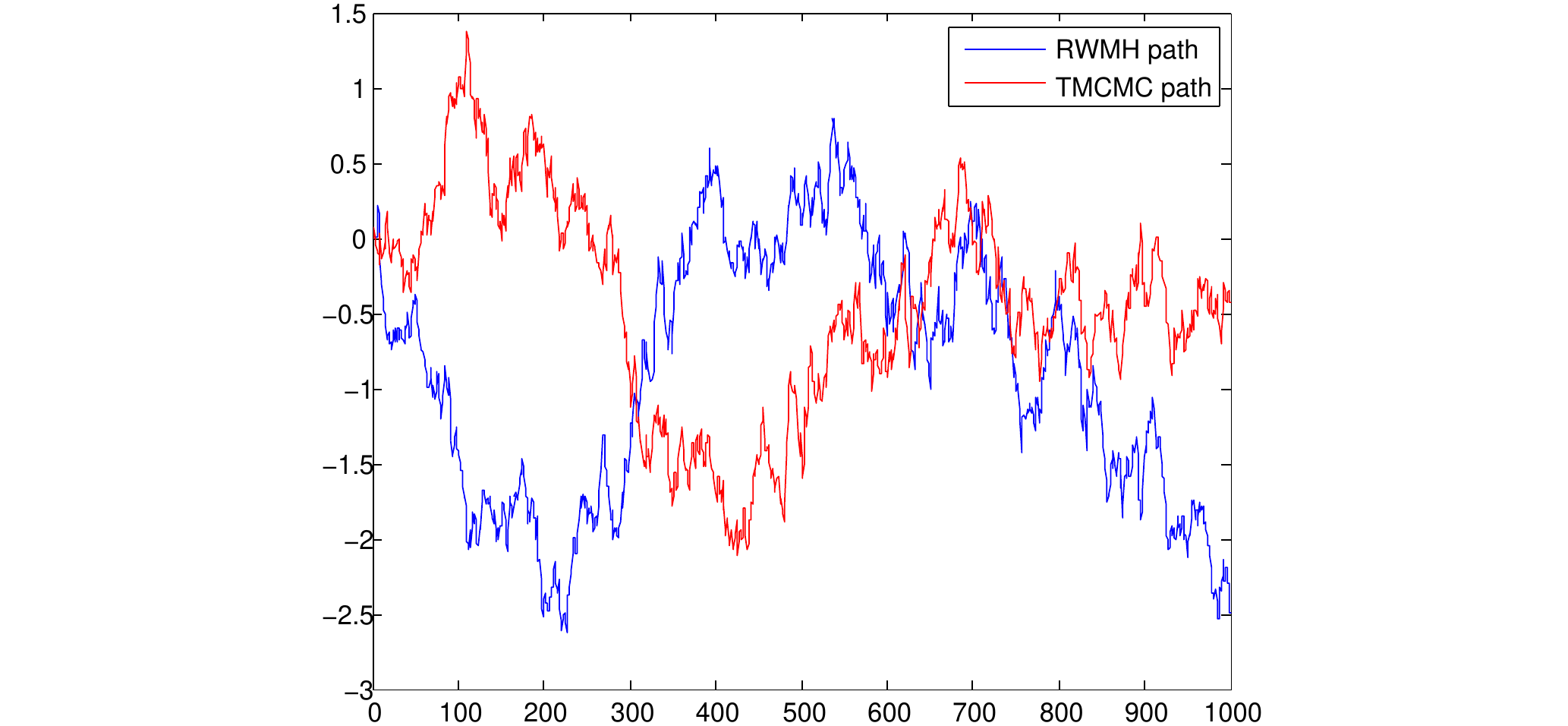}}\\
\subfigure [ Large proposal variance sample path]{ \label{fig:fig4}
\includegraphics[trim= 0cm 10cm 0cm 10cm, clip=true, width=7cm,height=3cm]{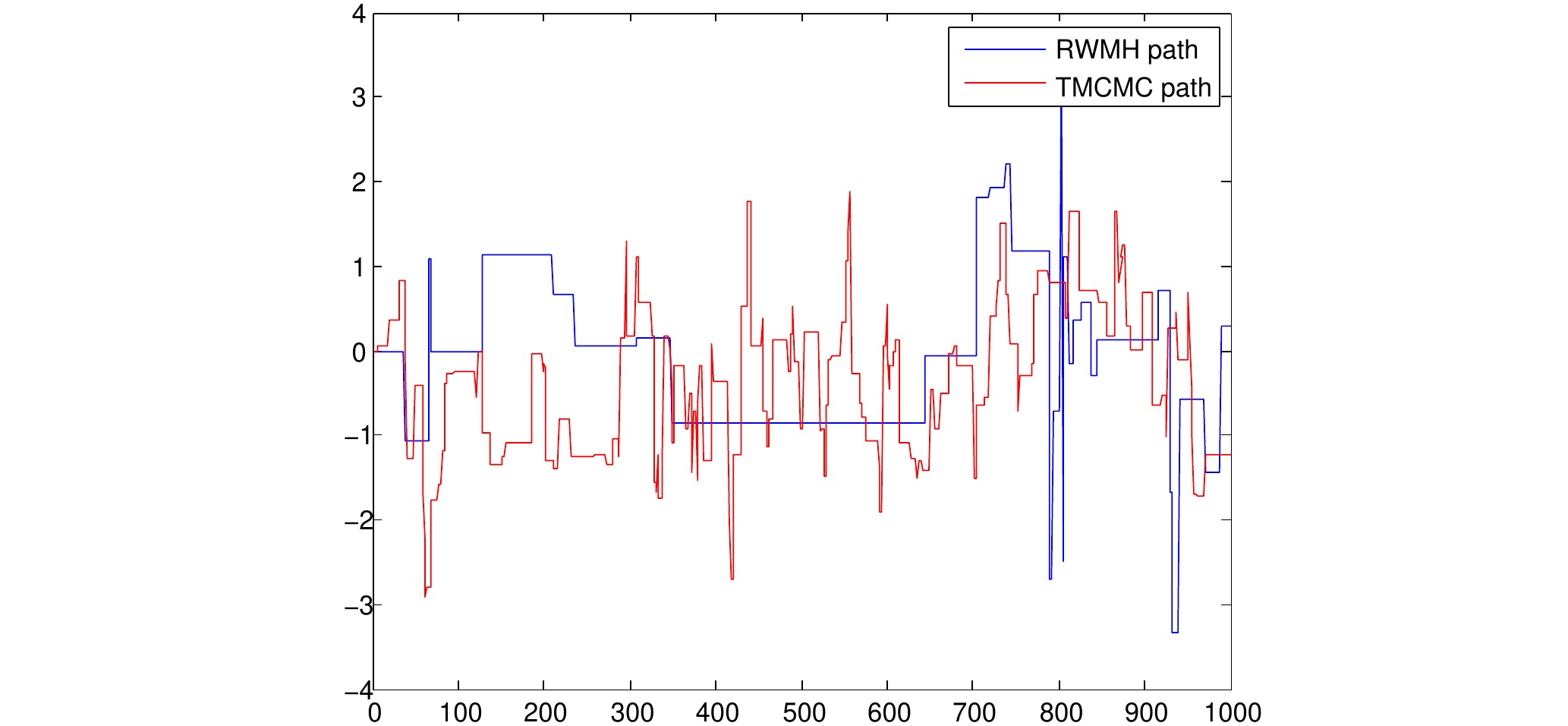}}
\caption{\emph{The graphical representation of a co-ordinate for a 5-dimensional chain with target density 
being product of $N(0,1)$ densities and the values of the scaling factor $l$ for the two cases are 
taken to be $l=0.8$ and $l=8$ respectively for the two scenarios a) and b) depicted in the graph}}
\label{fig:figoptscale}
\end{figure}

In this section, we shall restrict our focus on target densities that are products of 
iid components $ \pi =\prod_{i=1}^{d} f $ and the proposal density for $\epsilon $ is given by 
$TN_{>0} (0, \frac{l^2}{d})$, where $l$ is called the scaling term of the proposal. 
This section will be dedicated to obtaining the optimal value of this scaling $l$ and
determining the limiting expected acceptance rate of ATMCMC under the optimal scaling scenario. 
If the variance of the proposal density is very small, then the jumps will be of smaller magnitude and 
this would mean the Markov chain would take very many iterations to traverse the entire state space, 
and in the process, the convergence rate would be very small. On the other hand, if the variance 
is very large, then our algorithm will reject too many of the moves. 
An instance of this argument is depicted in \textbf{Fig~\ref{fig:figoptscale}}.

There is an extensive theory on optimal scaling of RWMH chains (see Beskos, Roberts and Stuart \cite{Beskos2009}, 
Bedard \cite{Bedard2009} \cite{Bedard2007}, Neal and Roberts \cite{NealRoberts}, Roberts, Gelman and Gilks \cite{Gelman}). 
The magic number for RWMH has been the optimal acceptance rate value of 0.234, which has been achieved 
through maximization of speed of the process for a wide range of distributions - iid set up, some 
special class of independent but non-identical set up, as well as a dependent set-up. 
For our purpose, we have developed an optimal scaling theory for ATMCMC where we 
have optimized the diffusion speed of our process to obtain optimal acceptance rate for ATMCMC. 
We present a rough sketch of our approach here, 
for detailed analysis we refer the reader to Dey and Bhattacharya \cite{Dey2013(2)}.

We assume that $f$ is Lipschitz continuous and satisfies the following conditions: 
\begin{equation}\label{eq:C1}
(C1)   \hspace{0.3 cm} E\left [ \left \{ \frac{f^{'}(X)}{f(X)} \right \}^{8} \right ] 
\hspace{0.1 cm}= \hspace{0.1 cm} M_{1} \hspace{0.1 cm} < \infty. 
\end{equation}

\begin{equation}\label{eq:C2}
(C2)   \hspace{0.3 cm} E\left [ \left \{ \frac{f^{''}(X)}{f(X)} \right \}^{4} \right ] 
\hspace{0.1 cm}= \hspace{0.1 cm} M_{2} \hspace{0.1 cm} < \infty. 
\end{equation}

We define ${U_{t}}^{d} = {X^{d}_{[dt],1}}$, the sped up first component of the actual Markov chain. 
Note that this process makes a transition at an interval of $\frac{1}{d}$. As we set $d \rightarrow \infty$, 
meaning that as the dimension of the space blows to $\infty$, the sped up ATMCMC process essentially 
converges to a continuous time diffusion process. 

For our purpose, we define the discrete time generator of the TMCMC approach, as 

\begin{eqnarray}\label{eq:generator}
G_{d}{V(x)}&=& \frac{d}{2^{d}} \displaystyle \sum_{ \left \{\begin{array}{l} b_{i}\in \{-1,+1\} \\  
\forall i=1,\ldots,d \end{array}\right \}} \displaystyle \int_{0}^{\infty} 
\left [\left ( \vphantom{min \left \{ 1, 
\frac{\pi(x_{1}+b_{1}\epsilon, \ldots, x_{d}+b_{d}\epsilon)}{\pi(x_{1},\ldots, x_{d})} \right \}} 
V \left (x_{1}+b_{1}\epsilon, \ldots, x_{d}+b_{d}\epsilon \right) - V \left (x_{1},\ldots, x_{d}\right ) \right) 
\right. \nonumber \\
&& \qquad \left. \hspace{3 cm} \times 
\left ( \normalsize \min \left \{ 1, \frac{\pi(x_{1}+b_{1}\epsilon, \ldots, x_{d}+b_{d}\epsilon)}{\pi(x_{1},x_{2}, 
\ldots, x_{d})} \right \} \right ) \right ] q(\epsilon)d\epsilon. \nonumber \\
\end{eqnarray}
In the above equation, we may assume that $V$ belongs to the space of inifinitely differentiable functions on compact support 
(see, for example, \ctn{Bedard2007})
for further details).

Note that this function is measurable with respect to the Skorokhod topology and we can treat $G_d$ 
as a continuous time generator that has jumps at the rate $d^{-1}$. Given our restricted focus on a 
one dimensional component of the actual process, we assume 
$V$ to be a function of the first co-ordinate only. Under this assumption, the generator defined in (\ref{eq:generator}) 
is a function of only $\epsilon$ and $b_{1}$, and can be rephrased as 

\begin{eqnarray}\label{eq:realgenerator}
G_{d}{V(x)} &=& \frac{d}{2}\int_{0}^{\infty}\sum_{b_{1}\in \{-1,+1\}} 
\left [\left (\vphantom{\min \left \{ 1, \frac{\pi(x_{1}+b_{1}\epsilon, \ldots, x_{d}+b_{d}\epsilon)}{\pi(x_{1},\ldots, x_{d})} \right \}} 
V(x_{1}+b_{1}\epsilon) - V (x_{1}) \right)  \right. \nonumber \\
&& \qquad \quad \left. \hspace{2 cm} \times E_{b_{2},\ldots, b_{d}}
\left ( \normalsize \min \left \{ 1, \frac{\pi(x_{1}+b_{1}\epsilon, \ldots, x_{d}+b_{d}\epsilon)}{\pi(x_{1},\ldots, x_{d})} \right \} \right ) 
\right ] q(\epsilon)d\epsilon, \nonumber \\
\end{eqnarray}
where $E_{b_{2},\ldots,b_{d}}$ is the expectation taken conditional on $b_{1}$ and $\epsilon$. 
%Note that if $\epsilon$ is taken to be truncated Normal $(0,1)$, with left truncation at $0$, then $b_{i}\epsilon$ 
%for each $i =1,2,\ldots, d $ follows $N(0,1)$ and since $E (b_{i}b_{j} \epsilon)$ for $i \neq j$ is also $0$, we can say that 
%$b_{i}\epsilon$ are uncorrelated. However, $b_{i}\epsilon$ are not independent because if they were, the pairwise sum would have been normal. 
%However, since $ b_{i}\epsilon+b_{j}\epsilon$ is equal to 0 with probability $\frac{1}{2}$ if $i \neq j$, the normality assumption for 
%the sum is contradicted. Also, if we are working on a $d$ dimensional space, we assume that 
%$\epsilon \sim TN^{L}_{0} (0, \frac{l^{2}}{d})$ where we use the notation $TN^{L}_{0}$ means truncated normal with left truncation at $0$.  

First we show that the quantity $G_{d}V(x)$ is a bounded quantity. 
\begin{eqnarray}\label{eq:bdd}
G_{d}{V(x)} &\leq & d E_{\{b_1,\epsilon\}}\left[ V(x_1+b_1\epsilon) - V(x_1) \right]  \nonumber \\
&=& dV^{'}(x_1)E_{\{b_1,\epsilon\}}(b_1\epsilon) + \frac{d}{2}V^{''}(x^*_1)E_{\{b_1,\epsilon\}}(\epsilon^{2}) \nonumber \\
&\leq&  l^{2}M_V, \nonumber  \\
\end{eqnarray}
where $x^*_1$ lies between $x_1$ and $x_1+b_1\epsilon$ and $M_V$ is the maximum value of $V^{''}$. %, provided it is bounded. 
%Most of the common choices of $V$ are bounded, enabling us to study its maximum value.

We derive the limit of $G_{d}V(x)$ as $d \rightarrow \infty$ that will give us the infinitesimal generator 
of the associated diffusion process for the ATMCMC chain. It can be shown that

 \begin{proposition}\label{prop1}

If $X \sim N(\mu,\sigma^{2})$, then 
\begin{equation}
E \left [ \min \left \{1, e^{X}\right \} \right] = 
\Phi \left (\frac{\mu}{\sigma}\right ) + e^{\left\{\mu +\frac{\sigma^{2}}{2}\right\}}
\Phi \left ( -\sigma - \frac{\mu}{\sigma} \right ), 
\end{equation}
where $\Phi$ is the standard Gaussian cdf. 

\end{proposition}

Using this proposition, we can write 

\begin{eqnarray}\label{eq:simpexp}
& E \bigg|_{b_{1}\epsilon} & \left 
[\min \left \{ 1, \frac{\pi(x_{1}+b_{1}\epsilon, \ldots, x_{d}+b_{d}\epsilon)}
{\pi(x_{1}, \ldots, x_{d})} \right \} \right ]\nonumber  \\
&=& \Phi \left (\frac{\eta (x_{1}, b_{1}, \epsilon) - \frac{(d-1)\epsilon^{2}}{2}\mathbb{I}}
{\sqrt{(d-1)\epsilon^{2}\mathbb{I}}}\right ) 
+ e^{\eta (x_{1}, b_{1}, \epsilon)}\Phi \left (-\sqrt{(d-1)\epsilon^{2}\mathbb{I}} 
- \frac{\eta (x_{1}, b_{1}, \epsilon) 
- \frac{(d-1)\epsilon^{2}}{2}\mathbb{I}}{\sqrt{(d-1)\epsilon^{2}\mathbb{I}}} \right) \nonumber\\
&=& \mathbb{W}(b_{1}\epsilon, x_{1}). \nonumber \\
\end{eqnarray}

Note that using Taylor series expansion around $x_{1}$, we can represent $\eta (x_{1}, b_{1}, \epsilon)$ as 
%write (\ref{eq:eta}) as 
\begin{equation}
\eta (x_{1}, b_{1}, \epsilon) = b_{1}\epsilon \left [\log f(x_{1}) \right ]^{'} + \frac{\epsilon^{2}}{2}
\left [\log f(x_{1}) \right] ^{''} + b_{1}\frac{\epsilon^{3}}{3!}\left [\log f(\xi_{1}) \right] ^{'''},
\end{equation}
where $\xi_1$ lies between $x_1$ and $x_1+b_1\epsilon$. 
Again re-writing $b_{1}\epsilon$ as $\frac{l}{\sqrt{d}}z^*_1$, 
where $z^*_1$ follows a $N(0,1)$ distribution, $\eta$ and $\mathbb W$ 
can be expressed in terms of $l$ and $z^*_1$ as 

\begin{equation}
\eta (x_{1}, z^*_1, d) = \frac{l z^*_{1}}{\sqrt{d}} \left [\log f(x_{1}) \right ]^{'} 
+ \frac{l^{2}{z^*_{1}}^{2}}{2!d}\left [\log f(x_{1}) \right] ^{''} + \frac{l^{3}{z^*_{1}}^{3}}
{3!d^{\frac{3}{2}}}\left [\log f(\xi_{1}) \right] ^{'''}
\end{equation}
and
\begin{equation}\label{eq:W}
\mathbb{W}(z^*_{1},x_{1},d) =  
\Phi \left (\frac{\eta (x_{1}, z^*_{1}, d) - \frac{{z^*_{1}}^{2}l^{2}}{2}\mathbb{I}}
{\sqrt{{z^*_{1}}^{2}l^{2}\mathbb{I}}}\right ) + e^{\eta (x_{1}, z^*_{1}, d )}
\Phi \left (\frac{-\frac{{z^*_{1}}^{2}l^{2}\mathbb{I}}{2} - \eta (x_{1}, z^*_{1}, d)}
{\sqrt{{z^*_{1}}^{2}l^{2}\mathbb{I}}} \right).
\end{equation}
The last line follows as the expression $ \eta (x_{1}, b_{1}, \epsilon)$ depends on $b_{1}$ 
and $\epsilon$ only through the product $b_{1}\epsilon$. 

Now we consider the Taylor series expansion around $x_{1}$ of the term 
\begin{eqnarray}
&& d E_{z^*_{1}} \left [\vphantom{\frac{1}{2}}\left ( V\left (x_{1}+ \frac{z^*_{1}l}{\sqrt{d}}\right ) 
- V \left (x_{1}\right ) \right) \mathbb{W} \left (z^*_{1}, x_{1},d\right ) \right] \nonumber \\
&=& d E_{z^*_{1}} \left [ \left \{ V^{'}(x_{1})\frac{z^*_{1}l}{\sqrt{d}} 
+ \frac{1}{2} V^{''}(x_{1})\frac{{z^*_{1}}^{2}l^{2}}{d} 
+ \frac{1}{6}V^{'''}(\xi_{1})\frac{{z^*_{1}}^{3}l^{3}}{d^{\frac{3}{2}}} \right \} 
\mathbb{W} \left (z^*_{1}, x_{1},d\right) \right ]. \nonumber \\
%&& X \left [ \mathbb{W}(0) + b_{1}\epsilon \mathbb{W}^{'}(0) + \frac{1}{2}\epsilon^{2}\mathbb{W}^{''}(\xi) \right] \nonumber \\
%&= & b_{1}\epsilon \left [  V^{'}(x_{1})\mathbb{W}(0) \right ] + \epsilon^{2}\left [ \frac{1}{2} V^{''}(x_{1}) \mathbb{W}(0)  + \frac{1}{2} \mathbb{W}^{'}(0) V^{'}(x_{1}) \right]  \nonumber \\
%&&+ b_{1}\epsilon^{3} \left [  \frac{1}{2} \mathbb{W}^{''}(\xi) V^{'}(x_{1}) + \frac{1}{6} \mathbb{W}(0) V^{'''}(z_{1}) \right] + \frac{1}{6}\epsilon^{4} \left [\mathbb{W}^{'}(0)V^{'''}(z_{1}) \right ] + \frac{1}{12} b_{1}\epsilon^{5}\left [ \mathbb{W}^{''}(\xi)V^{'''}(z_{1}) \right].\nonumber \\
\end{eqnarray}

From (\ref{eq:W}) it is clear that $\mathbb{W}(z^*_{1}, x_{1}, d)$ is continuous but not differentiable 
at the point $0$. Using Taylor series expansion of the terms $\Phi \left (\frac{\eta (x_{1}, z^*_{1}, d) - \frac{{z^*_{1}}^{2}l^{2}}{2}\mathbb{I}}
{\sqrt{{z^*_{1}}^{2}l^{2}\mathbb{I}}}\right )$, $e^{\eta (x_{1}, z^*_{1}, d )}$ and 
$\Phi \left (\frac{-\frac{{z^*_{1}}^{2}l^{2}\mathbb{I}}{2} - \eta (x_{1}, z^*_{1}, d)}
{\sqrt{{z^*_{1}}^{2}l^{2}\mathbb{I}}} \right)$ about $ \eta =0$, we obtain the expression of $G_{d}(V(x)$ as 

\begin{equation}
G_{d}{V(x)} = V^{'}(x_{1}) \frac{l^2}{2} [log f(x_1)]^{'} E_{z^*_{1}} 
\left [ {z^*_{1}}^2 \mathcal{V} \left (z^*_{1} \right)\right] 
+ \frac{1}{2}V^{''}(x_{1})l^{2}E_{z^*_{1}} \left [ {z^*_{1}}^{2}\mathcal{V} \left (z^*_{1} \right) 
+ \mathcal{O}(d^{-\frac{1}{2}})\right].
\end{equation}
where 
\begin{equation}
\mathcal{V} \left (z^*_{1} \right)  \rightarrow 2 \Phi \left (- \frac{|z^*_{1}|l\sqrt{\mathbb{I}}}{2}\right)
=2\left[1-\Phi \left (\frac{|z^*_{1}|l\sqrt{\mathbb{I}}}{2}\right)\right].
\end{equation}

The infinitesimal generator $GV(x)$ obtained as the limit of the $GV_d(x)$  has therefore a simpler form

\begin{eqnarray}
GV(x) = h(l) \left [ \frac{1}{2} (\log f)^{'}(x_{1})V^{'}(x_{1})+ \frac{1}{2}V^{''}(x_{1}) \right ].
\end{eqnarray}

This is the form of the generator for a Langevin diffusion process with 

\begin{equation}
h_{ATMCMC}(l) = 4l^{2} \int_{0}^{\infty} {z^{2} \Phi \left (- \frac{\sqrt{{z_{1}}^{2}l^{2}\mathbb{I}}}{2}\right)}.
\end{equation}

The function $h$ is called the diffusion speed and we maximize this quantity with respect to 
$l$ to derive the optimal scaling. For our case, $l_{opt} = \frac{2.4}{\sqrt{I}}$ and we plug this value 
in the formula for asymptotic expected acceptance rate to obtain 

\begin{equation}
%\alpha_{opt}=4\int_0^{\infty} \Phi\left (- \frac{\sqrt{{u}^{2}l^{2}_{opt}\mathbb{I}}}{2}\right)\phi(u)du.
\alpha_{opt}=4\int_0^{\infty} \Phi\left (- \frac{|u|l_{opt}\sqrt{\mathbb{I}}}{2}\right)\phi(u)du.
\label{eq:tmcmc_opt_acc}
\end{equation}

For RWMH too, the diffuion process is Langevin but the form of the diffusion speed is somewhat different 
(see Roberts, Gelman and Gilks \cite{Gelman}): 

\begin{equation}\label{eq:hl}
h_{RWMH}(l) = 2 l^{2} \Phi \left (\frac{-l \sqrt{I}}{2}\right ).
\end{equation}

\begin{figure}[h]
\centering
\includegraphics[width=10cm, height= 8 cm]{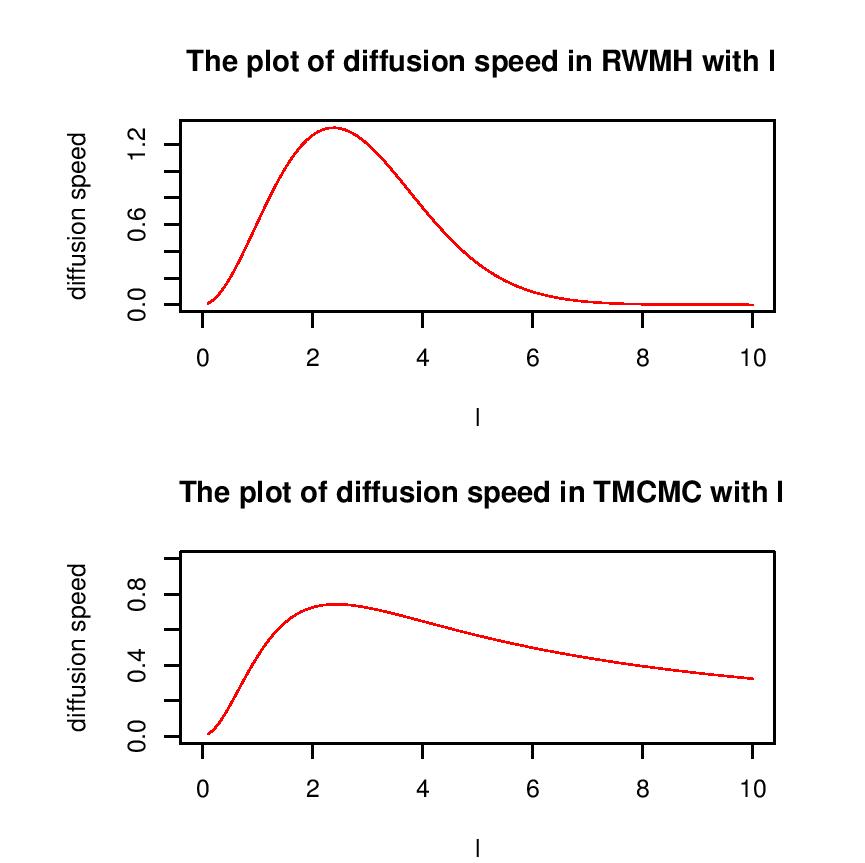}
\caption{\emph{The plot of the diffusion speed with respect to the scaling factor $l$ for RWMH and ATMCMC chains.}}
\label{fig:diffspeed}
\end{figure}

It was noted in \cite{Gelman} that the limiting expected acceptance rate corresponding to optimal scaling in 
RWMH is 0.234, while for that for the optimal scaling in additive TMCMC is $0.439$ which is almost twice as that of RWMH. 
It is to be noted that the optimal scale of RWMH is $l_{opt}=\frac{2.4}{\sqrt{I}}$, which, up to the first decimal place, 
is the same as that of ATMCMC. 
%This shows that indeed additive TMCMC has much higher expected acceptance rate compared to RWMH. 
The graphs of the diffusion speeds over different $l$ for ATMCMC and for standard RWMH are 
presented in \textbf{Fig~\ref{fig:diffspeed}}. \\[3 pt]

Note that the diffusion speed at $l_{opt}$ is higher for RWMH compared to additive TMCMC (ATMCMC)
implying that once stationarity is reached, there will be faster mixing among the iterates in 
RWMH compared to ATMCMC. However, an interesting observation is that if $l$ 
deviates slightly from $l_{opt}$, the diffusion speed of RWMH drops much faster compared to that of ATMCMC. 
Thus, ATMCMC is much more robust compared to RWMH with respect to the scaling. This is very important
in complex and high-dimensional practical situations where achieving the optimal scaling usually turns
out to be infeasible; recall the discussion regarding this in Section 1. Although our above analysis holds true only for the case when all the components of the product density are \emph{iid}, however, this condition can be relaxed to include independent components with appropriate scaling and inherent regularization properties as in Bedard (2009) \cite{Bedard2009} and Dey and Bhattacharya (2013) \cite{Dey2013} and also to  non-regular component densities in Dey and Bhattacharya \cite{Dey2014}.

Also, in all the calculations we have done so far and in the consideration of the diffusion speed 
and its implications, we must keep in mind our inherent assumption that the process is in stationarity. 
The major question to address now is that which chain has faster convergence to stationarity. 
We address this in the next section via simulation studies.
%for both the chains and measuring the correspondence of the empirical distributions at each time point 
%for RWMH and additive TMCMC with respect to that of the target density.

\section{Simulation study comparison}

In this section, we compare RWMH and additive TMCMC methods using two parameters, 
one being the acceptance rate and the other, the Kolmogorov-Smirnov (KS) distance  
between the empirical distribution at each time point and the target density. For the first measure, we observed 
the acceptance rates of the two algorithms for varying dimesnions and scaling factors $l$. 
The results are reported in \textbf{Table~\ref{tab:tab1}}. \\[2 pt]

\begin{table}[h]

\centering
\resizebox{9 cm}{4 cm}{
\begin{tabular}{|p{0.4in}|c|c|c|}
\hline
\multirow{2}{*}{Dim} & \multirow{2}{*}{\backslashbox{Scaling}{Test}} & \multicolumn{2}{|c|}{$\begin{array}{c} Acceptance \\ rate ($\%$) \end{array} $} \\ \cline{3-4}
& & RWMH & TMCMC\\ \hline

\multirow{3}{*}{2} & 2.4 & 34.9 & 44.6  \\ 
 & 6 & 18.66 & 29.15 \\ 
& 10 & 3.83 & 12.36 \\ \hline
\multirow{3}{*}{5} & 2.4 (opt) & 28.6 & 44.12  \\ 
& 6 & 2.77 & 20.20 \\ 
& 10 & 0.45 & 12.44\\ \hline
\multirow{3}{*}{10} & 2.4 (opt) & 25.6 & 44.18  \\ 
& 6 & 1.37 & 20.34 \\ 
& 10 & 0.03 & 7.94\\ \hline
\multirow{2}{*}{100} & 2.4 (opt) & 23.3 & 44.1 \\ 
& 6 & 0.32 & 20.6  \\ \hline
\multirow{2}{*}{200} & 2.4 (opt) & 23.4 & 44.2  \\ 
& 6 & 0.33 & 20.7 \\ \hline
\end{tabular} }
\caption{\emph{Table representing the acceptance rates of RWMH and ATMCMC approaches for varying
dimensions and varying scaling factors $l$, with the target density given by a iid product of $N(0,1)$ densities.}}
\label{tab:tab1}
\end{table}

\textbf{Table~\ref{tab:tab1}} validates that for higher dimensions, under optimal scaling, 
the acceptance rates of RWMH and additive TMCMC are indeed $0.234$ and $0.439$ respectively, 
as the observed values are very close to the theoretical ones. Also, we see that for fixed dimensions, 
as scaling increases away from the optimal value, the acceptance rate falls drastically 
for RWMH and this worsens with increase in dimensionality. 
For dimensions $100$ and $200$, we skipped providing the acceptance rates for scaling $l=10$ 
as it was understandably very small for RWMH. Comparatively, additive TMCMC is much more stable 
with change of scaling even for high dimensions. This validates the robustness of the diffusion 
speed with respect to scaling $l$ in \textbf{Fig~\ref{fig:diffspeed}}. \\[3 pt]

\begin{figure}%[htp]
\centering
\subfigure [$d=30,~l=2.4.$]{ \label{fig:KS1_30}
\includegraphics[trim= 0cm 8cm 0cm 8cm, clip=true, width=7cm,height=4cm]{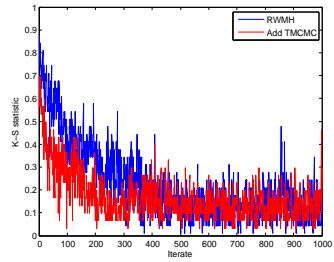}} \hspace{-1 cm}
\subfigure [$d=30,~l=4.$]{ \label{fig:KS2_30}
\includegraphics[trim= 0cm 8cm 0cm 8cm, clip=true, width=7cm,height=4cm]{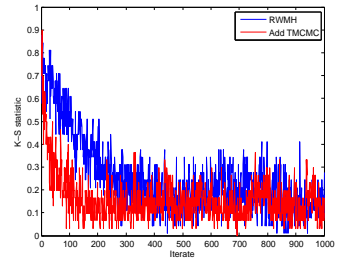}}
\caption{\emph{The KS distance graph for RWMH and ATMCMC chains for a $30$ dimensional target density, which is
the product of iid $N(0,1)$ components. The scalings for the two graphs are $l=2.4$ and $l=4$. Notice that the 
KS graph for ATMCMC seems to be lower compared to that of RWMH implying faster rate of convergence for  ATMCMC.}}
\label{fig:ks1}
\end{figure}

\begin{figure}[h]
\subfigure [$d=100, ~l=2.4.$]{ \label{fig:KS1_100}
\includegraphics[trim= 0cm 8cm 0cm 8cm, clip=true, width=7cm,height=4cm]{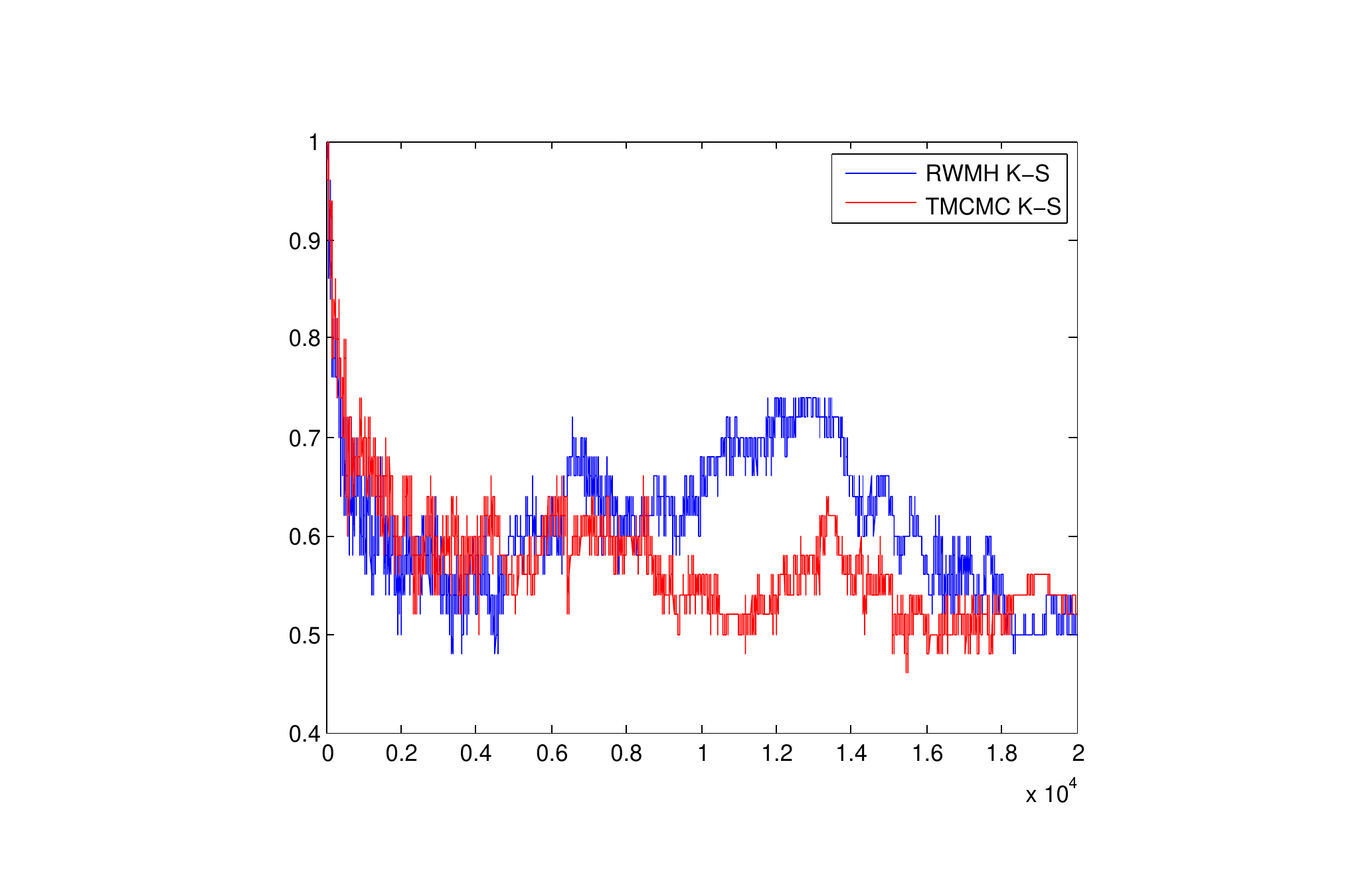}} 
\hspace{-1cm}
\subfigure [$d=100, ~l=4.$]{ \label{fig:KS2_100}
\includegraphics[trim= 0cm 8cm 0cm 8cm, clip=true, width=7cm,height=4cm]{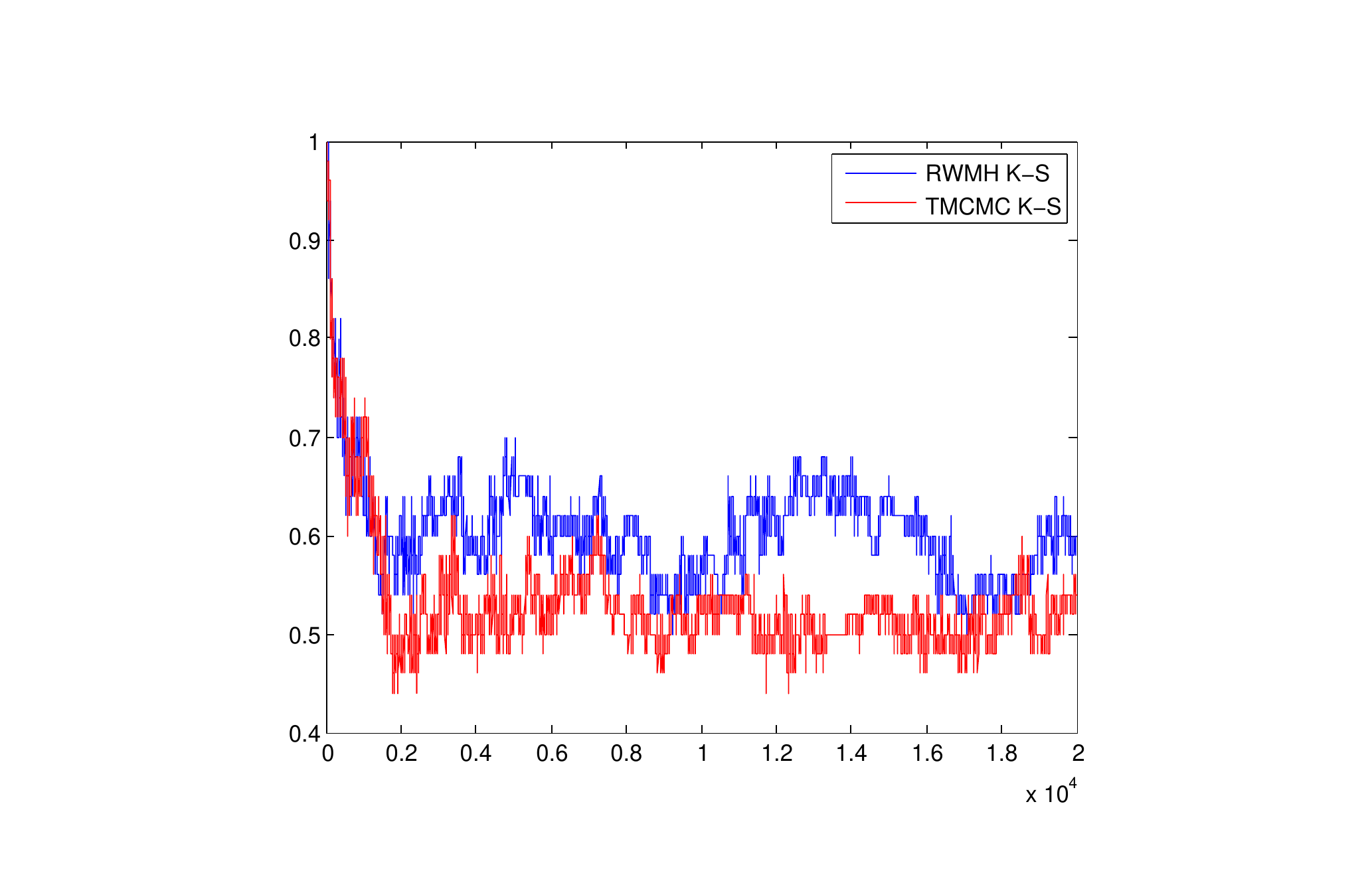}}
\caption{\emph{The KS distance graph for RWMH and ATMCMC chains for a $100$ dimensional target density, which is
the product of iid $N(0,1)$ components. The scalings for the two graphs are $l=2.4$ and $l=4$. Here the 
KS graph for ATMCMC is clearly lower compared to that of RWMH implying faster rate of convergence for ATMCMC.}}
 \label{fig:ks2}
\end{figure}

For the second measure of KS distance comparison, we run a number of chains, say L, 
starting from one fixed point for both RWMH and ATMCMC adaptations. Corresponding to each time point $t$, 
we thus get L many iterates. The notion is that, as time $t$ increases (specially after burn-in), 
these L many iterates should be close to an independently drawn random sample from the target distribution $\pi$. 
So, if we observe the KS statistic for the empirical distribution of these iterates along any particular 
dimension with respect to the marginal of $\pi$ along that dimension, we expect the test statistic to be 
decreasing with time and finally being very close to 0 after a certain time point. Now the question of interest 
is, of the two approaches, ATMCMC and RWMH, for which method the graph decays faster to $0$? 
Corresponding to two different dimensions $d=10$ and $d=100$, and two scalings $l=2.4$ 
(optimal given that $\mathbb{I}=1$ for the target density product of $N(0,1)$ components) and $l=4$, 
we present the two graphs of additive TMCMC and RWMH simultaneously in 
\textbf{Fig~\ref{fig:ks1}} and \textbf{Fig~\ref{fig:ks2}}. Both the figures, but particularly the latter,
clearly indicate faster convergence of ATMCMC to the stationary distribution.
\\[2pt]

Therefore in conclusion it can be stated that

\begin{itemize}
\item ATMCMC is simple to interpret and does not depend heavily on the target density, 
and additionally has much lesser computational burden and time complexity.
\vspace{0.2 cm}
\item Under sub-exponential target density with some regularity constraints on the target density, 
the ATMCMC algorithm is geometrically ergodic.
\vspace{0.2 cm}
\item ATMCMC has a higher acceptance rate of 0.439 corresponding to 0.234 for the RWMH algorithm. 
As observed, our algorithm is more robust to change of scale and across dimensions. But the mixing 
or diffusion speed of RWMH is higher, meaning that once stationarity is attained RWMH 
will provide better samples than ATMCMC.
\vspace{0.2 cm}
\item The KS test comparison in the simulation study shows that for high dimensions, ATMCMC has 
lower KS statistic value compared to RWMH when the chain is not stationary. This also suggests that 
ATMCMC reaches burn-in faster than RWMH for higher dimensions. But once burn-in is reached, 
ideally the two methods should both yield KS values close to $0$ and that is why we see that the KS graphs 
stabilize with time for both the approaches.

\end{itemize}

\newpage

%We assume that $\log (f)$ is twice continuously differentiable and 
%$\left[ \log (f)\right]^{(k)}$ is bounded for $k =1(|)3$. 
%Under this assumption, we can neglect the cubic terms of $z_{1}$, and hence essentially our generator reduces to 

% Bibliography and Glossary          (\phantomsection is needed for hyperlinks)

%\phantomsection%
%\addcontentsline{toc}{chapter}{\bibname}              % add Bibliography to TOC
%\bibliographystyle{alpha}\bibliography{mcmc}

\end{document}